# Current jetting distorted planar Hall effect in a Weyl semimetal with ultrahigh mobility

J. Yang,[1,2] W. L. Zhen,[1] D. D. Liang,[1] Y. J. Wang,[1] X. Yan,[1] S. R. Weng,[1] J. R. Wang,[1] W. Tong,[1] L. Pi,[1,2,*] W. K. Zhu,[1,†] and C. J. Zhang[1,3,‡]

[1]*Anhui Province Key Laboratory of Condensed Matter Physics at Extreme Conditions, High Magnetic Field Laboratory, Chinese Academy of Sciences, Hefei 230031, China*

[2]*Hefei National Laboratory for Physical Sciences at Microscale, University of Science and Technology of China, Hefei 230026, China*

[3]*Institute of Physical Science and Information Technology, Anhui University, Hefei 230601, China*

A giant planar Hall effect (PHE) and anisotropic magnetoresistance (AMR) is observed in TaP, a nonmagnetic Weyl semimetal with ultrahigh mobility. The perpendicular resistivity (i.e., the planar magnetic field applied normal to the current) far exceeds the zero-field resistivity, which thus rules out the possible origin of negative longitudinal magnetoresistance. The giant PHE/AMR is finally attributed to the large anisotropic orbital magnetoresistance that stems from the ultrahigh mobility. Furthermore, the mobility-enhanced current jetting effects are found to strongly deform the line shape of the curves, and their evolution with the changing magnetic field and temperature is also studied. Although the giant PHE/AMR suggests promising applications in spintronics, the enhanced current jetting shows the other side of the coin, which needs to be considered in the future device design.

# I. INTRODUCTION

Topological (Dirac and Weyl) semimetals (TSMs) attract fast growing interests for their intriguing properties, such as extremely large non-saturating positive magnetoresistance (XMR) [1], chiral anomaly induced negative longitudinal MR (NLMR) [2], surface Fermi arcs [3,4], etc. Chiral anomaly refers to the non-conservation of chiral charge around the Weyl nodes when the applied electric and magnetic fields are non-orthogonal ($E \cdot B \neq 0$). As a result of the chiral anomaly, the NLMR has been widely used to investigate and identify the TSMs [5-10]. However, recent studies indicate that the measurement of NLMR can be affected by some extrinsic effects, like ionic impurity induced scattering [11], weak localization [12], conductivity fluctuation [13], and current jetting effect [14,15]. Especially for the semimetals with high mobility (e.g., TaP family), considerable concerns have been raised about the validity of NLMR, because the significantly enhanced current jetting effect can also induce a large NLMR [14].

Therefore, some new techniques are tried to unveil the nontrivial nature of TSMs. These techniques include the measurements of anomalous Hall effect [16], anomalous Nernst effect [17,18], and nonlinear optical response [19]. Notably, a recently arising phenomenon, i.e., the giant planar Hall effect (PHE), is intensively studied both theoretically [20,21] and experimentally [22-27], for its possible connection with the chiral anomaly in TSMs [22-24]. However, some studies show that the chiral anomaly induced PHE is just the angular dependence of NLMR [22,23]. The PHE measurements suffer from all the extrinsic effects that affect the measurements of NLMR, such as current jetting. Besides chiral anomaly, PHE may have other origins in various systems, including anisotropic magnetic scattering [28,29], topological surface state [30], and orbital magnetoresistance [27]. The different mechanisms are reflected as the contribution to the anisotropic resistivity $\Delta\rho = \rho_\perp - \rho_\parallel$, where $\rho_\perp$ and $\rho_\parallel$ are the resistivity corresponding to the (planar) magnetic field perpendicular to and parallel to the current (*I*), respectively. The PHE ($\rho_{yx}$) and related anisotropic MR (AMR, $\rho_{xx}$) can be expressed as [20]

$$\rho_{yx} = -\Delta\rho\sin\theta\cos\theta, \quad (1)$$

$$\rho_{xx} = \rho_\perp - \Delta\rho\cos^2\theta, \quad (2)$$

where $\theta$ is the angle of magnetic field with respect to current. Note that in Ref. [20] the formulations are based on the pure Weyl physics, where the $\Delta\rho$ comes only from the reduction of $\rho_\parallel$ (i.e., the NLMR) while the $\rho_\perp$ keeps unchanged (i.e., the zero-field resistivity $\rho_0$). However, in our following discussions the $\Delta\rho$ has contributions from all possible origins.

In this paper, we report the observation of giant PHE and AMR in a Weyl semimetal TaP, which is not naively associated with the chiral anomaly, because the ultrahigh mobility (~$10^6$ cm$^2$/Vs [15]) prohibits the observation of a chiral anomaly induced NLMR [14,24] and the chiral anomaly may be even absent in TaP [15]. The giant PHE/AMR is finally attributed to the large anisotropic orbital magnetoresistance (or XMR). In addition, the mobility-enhanced current jetting effects are found to strongly deform the line shape of the curves, and their evolution with the changing magnetic field and temperature is also studied. When the current jetting effects are suppressed, in terms of the low field and reduced mobility (by increasing temperature), well-defined PHE and AMR curves are obtained. Since the XMR is always related with the ultrahigh mobility [31-33], the giant PHE/AMR and the large current jetting in TaP are understood in the same scenario. Although the former suggests promising applications in spintronics, the latter shows the other side of the coin.

## II. EXPERIMENTAL METHODS

High quality TaP single crystals were synthesized using a chemical vapor transport method. Stoichiometric mixture of Ta and P powder were first heated in an evacuated fused silica tube for 48 hours at 850 ℃, and then the resultant polycrystals were sealed in a quartz tube with iodine as transport agent (9 mg/cm$^3$). Plate-like single crystals can be obtained after vapor transport growth with a temperature gradient from 930 ℃ to 820 ℃. The crystal structure and chemical composition were checked by single crystal x-ray diffraction (XRD) on a Rigaku-TTR3 x-ray diffractometer using Cu Kα radiation and on an Oxford Swift 3000 energy dispersive spectrometer (EDS). Magnetic susceptibility measurements were performed on a Quantum Design MPMS-3. Single crystals were polished until thin enough (around 150 um) for transport measurements that were taken on a Quantum Design PPMS. Standard four-probe technique was used to measure the longitudinal resistivity and Hall contacts were located on the transverse sides. Magnetic field was applied and rotated within the sample plane.

## III. RESULTS AND DISCUSSION

### A. Sample characterizations

The as-grown TaP single crystals are thin plates and exhibit square-like morphology, with a typical size of $3\times3\times0.8$ mm$^3$ [Fig. 1(a)]. The chemical composition is confirmed by EDS, as depicted in Fig. 1(b). Figure 1(c) shows the single crystal XRD pattern, in which two sharp single crystal peaks are detected. The high quality of single crystal is supported by the narrow full width at half maximum (FWHM, 0.08 °) revealed in the rocking curve scan [right inset of Fig. 1(c)]. The observed peaks are in good agreement with the (00$l$) diffraction of TaP with space group $I4_1md$ (No. 109), whose crystal structure is illustrated in the left inset of Fig. 1(c). The indices also suggest that the naturally cleaved surface is the $ab$ plane.

The obtained TaP single crystal is further characterized by the de Haas-van Alphen (dHvA) measurements and Fermi surface analyses. Figure 2(a) presents the magnetization as a function of magnetic field ($B \parallel c$) taken at various temperatures from 1.8 K to 10 K. On a diamagnetic background, the dHvA oscillations are superimposed, starting from a field as low as 0.6 T at 1.8 K. After removing the background (represented by a polynomial), the oscillations become more pronounced (plotted against 1/$B$) [Fig. 2(b)]. By performing fast Fourier transformation (FFT), the frequencies of oscillations and their harmonics are retrieved, i.e., 18 T ($\beta$), 24 T ($\gamma$), 29 T ($2\alpha$) and 46 T ($\delta$) [Fig. 2(c)]. These frequencies are highly consistent with previous Shubnikov-de Haas (SdH) and dHvA measurements [15,34,35], and are finally identified according to those results and band structure calculations [15].

The nature of Weyl fermions participating in quantum oscillations (e.g., high mobility) can be revealed by quantitative analyses of dHvA oscillations. The oscillations are described by Lifshitz-Kosevich (LK) formula [36], with the amplitude proportional to the thermal damping factor $R(T) = (\lambda m^*T/B)/\sin(\lambda m^*T/B)$, where $\lambda = 2\pi^2 k_B/e\hbar$, $k_B$ is the Boltzmann constant, $\hbar$ is the reduced Planck constant, and $m^*$ is the effective mass of carrier. Figure 2(d) shows the amplitudes as a function of temperature for all the frequencies. The fits to $R(T)$ yield effective mass for the corresponding carriers, i.e., $m_\alpha^* = 0.068\ m_0$, $m_\beta^* = 0.050\ m_0$, $m_\gamma^* = 0.071\ m_0$ and $m_\delta^* = 0.1\ m_0$ ($m_0$ is the mass of free electron). Such small effective mass is always linked to steep linear bands and ultrahigh mobility. Taking the $F_\beta$ branch for example, its extremal cross section area $A_F$ is calculated as 0.172 nm$^{-2}$, according to Onsager relation $F = (\hbar/2\pi e)A_F$. Supposing a circular cross section, a very small Fermi momentum $k_F$=0.234 nm$^{-1}$ is obtained, which further leads to a large Fermi velocity $v_F = \hbar k_F/m^* = 5.4 \times 10^5$ ms$^{-1}$. Such a large Fermi velocity is comparable with that of NbP which has an ultrahigh mobility [14], and also confirms the ultrahigh mobility in our TaP sample.

### B. Giant planar Hall effect

The measurement geometry of PHE and AMR is illustrated in the inset of Fig. 3(a). The standard four-probe technique is adopted to measure longitudinal MR ($\rho_{xx}$), along with two Hall contacts to measure planar Hall resistivity ($\rho_{yx}$). Magnetic field is applied within the sample plane (*ab* plane) and rotates around the *c* axis, with an angle $\theta$ relative to the current direction. In actual experimental set-up, the magnetic field does not always perfectly lie in the sample plane. That is, a small out-of-plane field component may exist (thus a small deviation angle $\varphi$), which will result in a regular Hall resistivity. One way to eliminate this term is to average the $\rho_{yx}$ data in positive and negative fields. Figure 3(a) shows the angular dependence of $\rho_{yx}$ taken at 2 K under a field of 2 T, after the positive/negative operation. Another two extrinsic effects may also exist due to the possible nonsymmetrical Hall contacts. The misaligned Hall contacts will induce a small longitudinal MR in the measured Hall resistivity. One part of the additional longitudinal MR is caused by the in-plane field, which has a $\cos^2\theta$ dependence according to Eq. (2), and another part is caused by the out-of-plane field, with an approximately $\sin^2(\theta+\varphi)$ dependence. Both terms cannot be removed by data processing. However, these two effects seem to be negligible in our measurements, as both of them are symmetrical for $\pm\theta$ which is distinctly different from the odd-function feature of the PHE curve in Fig. 3 (a).

The angular dependence of $\rho_{yx}$ has a period of $\pi$ and reaches its maximums at $\pm 45°$ and $\pm 135°$, both of which are consistent with the planar Hall effect. As shown in Fig. 3(a), the experimental data can be well fitted to Eq. (1), resulting in an anisotropic resistivity $\Delta\rho = 0.201$ mΩ cm. This value is comparable with that of GdPtBi [22,24] and Na$_3$Bi [24], and about one magnitude order larger than that of ZrTe$_5$ [26] and WTe$_2$ [25]. Figure 3(b) presents the angular dependence of $\rho_{xx}$, exhibiting a large planar AMR. Also, the AMR data can be fitted to Eq. (2). The fitting is roughly acceptable, except the misfit around 90° and 270°, which will be discussed in the next subsection. Here we focus on the possible origins of the giant PHE and AMR in TaP.

Within our knowledge scope, there are at least four possible origins of PHE/AMR in different systems. (i) In ferromagnetic metals, the AMR and PHE originate from the interplay of the magnetic order and the spin-orbit interactions. Obviously, the PHE in TaP cannot be attributed to this mechanism, taking account of its nonmagnetic nature. (ii) In topological insulators, the PHE and AMR arise from the topological protection mechanism through topological surface state, in which the spin orientation and momentum are locked. The backscattering would be significantly enhanced by the magnetic field that is parallel to the current (i.e., $B \perp$ spin orientation) and thus gives rise to an increased $\rho_\parallel$ [30]. Such a condition leads to a negative $\Delta\rho$, which is hence unlikely to account for the positive $\Delta\rho$ observed in TaP. (iii) In topological semimetals, the chiral anomaly induced NLMR is supposed to contribute to the PHE, namely, in terms of the reduced $\rho_\parallel$ when magnetic field is increased. However, as stated in the Introduction, the NLMR observed in TaP can be hardly related to the chiral anomaly, due to the enhanced current jetting effect by the ultrahigh mobility. Some band structure calculations even show that the chiral anomaly may be absent in TaP, as the Fermi surface connects pairs of Weyl points [15]. (iv) Orbital magnetoresistance arises from the asymmetric Fermi surface [37], which could in principle exist in any materials. The anisotropy of orbital magnetoresistance ($\rho_\perp > \rho_\parallel$) will definitely induce a positive $\Delta\rho$. Nevertheless, this effect is small in most cases. The giant AMR in TaP ($\rho_\perp \gg \rho_\parallel$) is attributed to the large anisotropic orbital magnetoresistance, as evidenced by the analyses below.

Figure 4(a) shows the angular dependence of $\rho_{yx}$ taken at 2 K and various magnetic fields. Although the increasing field strongly deforms the line shape which crosses over from a sine-type function to a nearly straight line, the essential features of PHE (i.e., period in $\pi$ and maximums at $\pm 45\,°$ and $\pm 135\,°$) remain. Rough fittings to Eq. (1) give the anisotropic resistivity $\Delta\rho$. We may note that the $\Delta\rho$ reaches nearly 5 m$\Omega$ cm at 14 T [inset of Fig. 4(a)], which is much larger than the $\rho_\parallel(B=0)$ (~u$\Omega$ cm). That is, the giant $\Delta\rho$ is unlikely to be attributed to the NLMR (trivial or nontrivial). As is further shown in the $B$-dependent AMR curves [Fig. 4(b)], the $\rho_\perp$ sharply rises while the $\rho_\parallel$ changes much less. The large resistivity anisotropy in response to the magnetic field is the origin of the giant $\Delta\rho$. As is known, TaP is a typical XMR semimetal [34]. More and more studies suggest that the most important origin of XMR is the ultrahigh mobility that is always related with the steep linear bands [31-33]. This can be qualitatively understood in the regime of two-band model, in which $\rho_\perp = [\frac{\sigma_e}{1+\mu_e^2 B^2} + \frac{\sigma_h}{1+\mu_h^2 B^2}]^{-1}$ and $\rho_\parallel = [\sigma_e + \sigma_h]^{-1}$ [24]. $\sigma_e = ne\mu_e$ and $\sigma_h = pe\mu_h$ are the electron and hole conductivity respectively, where $n$ ($p$) and $\mu_e$ ($\mu_h$) are the carrier density and mobility for electron (hole) carriers. For high $B$, we may deduce that $\Delta\rho \sim \frac{\mu_e^2 \mu_h^2 B^2}{\sigma_e \mu_h^2 + \sigma_h \mu_e^2}$. Namely, the ultrahigh mobility can solely induce a large $\Delta\rho$. Here for TaP, the quickly increasing $\rho_\perp$ with magnetic field and the resistivity plateau at low temperatures (unusually rising magnetoresistance at low temperatures, as shown in Fig. 4(d)) are consistent with the features of the XMR in TaP [34]. Hence, we finally attribute the giant PHE/AMR in TaP to the large anisotropic orbital magnetoresistance.

### C. Current jetting effects

The misfit of planar AMR in Fig. 3(b) is represented by a shoulder-like or double-peak structure around 90 ° (and 270 °), where it is supposed to be a single peak at exactly 90 ° (and 270 °). Similar structure has been observed in GdPtBi [22,24] and Bi and Sb [38]. We attribute it to the non-uniform current distribution in presence of magnetic field. As elucidated by A. B. Pippard [37], when *B* lies normal to a matchstick-shape sample, the current distribution and equipotentials are determined by the Hall angle if the sample has a Hall effect. That is, the equipotentials lie at the Hall angle to the normal of current flow, i.e., setting up an electric field along the Hall angle. The Hall angle is defined as $\tan^{-1}\omega_c\tau = \tan^{-1}\mu B$, where $\omega_c = \frac{eB}{m^*}$ is the cyclotron frequency, $\mu = \frac{e\tau}{m^*}$ is the mobility and $\tau$ is the relaxation time. The Hall angle may be close to 90 ° when $\mu B \gg 1$. For appropriate $\mu B$, the distribution of equipotentials may exhibit double peaks near 90 ° and a dip at 90 °. We note that this is one of current jetting effects, i.e., field induced anisotropic mobility of conductivity (the drift of carriers normal to *B* is suppressed compared to the drift along *B*). Current jetting effect is strongly dependent upon mobility and magnetic field, and is probably unavoidable for comparably large $\mu B$. Since various current jetting effects are discussed in the present paper, we classify the $B \perp I$ case as the *Hall-angle-type* current jetting effect, and the $B \parallel I$ case as the *classical* current jetting effect.

We further note that the PHE and AMR curves are strongly deformed by the enhanced current jetting effect [Figs. 4(a) and 4(b)], via the increasing magnetic field. As discussed above, current jetting effects depend on $\mu B$. Hence, for materials with high mobility, current jetting effect could be prominent at a relatively low field. If we define $B_c$ (inversely proportional to $\mu$) as the onset field, above which current jetting effects become evident [24], the $B_c$ of TaP family would be very low (~0.4 T) because of their ultrahigh mobilities (e.g., $2.5 \times 10^6$ cm$^2$/Vs for TaP and $5 \times 10^6$ cm$^2$/Vs for NbP [14,15]). That is, current jetting effects are usually inevitable in these materials. One disaster of current jetting effect is the harm to the identification of chiral anomaly by NLMR measurements, in which a chiral anomaly induced NLMR is expected when *B* is applied along or slightly misaligned with the current. However, this is hardly accessed for high-mobility materials as current jetting would also cause a large NLMR [14,15]. Another extrinsic phenomenon arising from current jetting effect is the so-called negative resistivity. Such a peculiar phenomenon has been observed in bismuth (with ultrahigh mobility) [38] and NbP [14] when $B$ is approaching the direction of current flow. As shown in Fig. 4(b), a negative resistivity also appears near 0° and 180°, and becomes more prominent as $B$ increases. Finally, the negative resistivity is attributed to the classical current jetting effect, and the deformation of $\rho_{yx}$ and $\rho_{xx}$ curves are caused by the combination of classical and Hall-angle-type current jetting effect.

Since the onset field of current jetting effect in TaP is very low, one feasible approach to suppress the current jetting effect is to decrease the mobility by increasing temperature. Due to the enhanced thermal fluctuation, the mobility of TaP is indeed sharply reduced (e.g., from $10^6$ cm$^2$/Vs at 2 K down to $10^3$ cm$^2$/Vs at 150 K) [34,39]. Figures 4(c) and 4(d) present the angular dependence of $\rho_{yx}$ and $\rho_{xx}$ taken at various temperatures and 14 T, respectively. It is found that, with the increasing temperature, the deformation is reduced and the curves accord with the theoretical formulas better. As shown in Fig. 5, the $\rho_{yx}$ and $\rho_{xx}$ curves taken at 250 K can be fitted fairly well by Eqs. (1) and (2), respectively, showing a suppressed current jetting effect. This can be understood in terms of the quickly rising $B_c$, as a result of the reduced mobility. The $B_c$ of TaP should far exceed 14 T at 250 K, if we consider its mobility is almost that of Na$_3$Bi or GdPtBi (3000 and 2000 cm$^2$/Vs at 2 K, respectively), whose $B_c$ can reach as high as 30 T [24]. Besides, the $\Delta\rho$ decreases quickly with the increasing temperature [inset of Fig. 4(c)], because of the reduced mobility. This is consistent with our explanation for the origin of PHE.

From Figs. 4(b) and 4(d), we can also find the evolution of Hall-angle-type current jetting effect with the changing magnetic field and temperature, as indicated by the red dashed lines. This kind of non-uniform current distribution is represented by the focusing of current flow along the direction of Hall angle, i.e., $\tan^{-1}\mu B$, when $B \perp I$. For the case with a moderate Hall angle (~70°-80° in our measurement), the double-peak structure appears and a dip forms at 90°. If the Hall angle is approaching 90° (i.e., $\mu B \gg 1$), the double peaks may merge into a single one. Therefore, the distance of double peaks ($\Delta\theta$) relies on the scale of $\mu B$. No matter which is increased ($\mu$ or $B$), the distance will decrease, even to zero for sufficiently large $\mu B$. This explains the occurrence and evolution of double-peak structure with the changing magnetic field and temperature. We note that similar phenomenon has been observed in GdPtBi [22,24]. Although the classical current jetting effect is unobservable in GdPtBi, the Hall-angle-type current jetting effect remains.

## IV. CONCLUDING REMARKS

The NLMR induced by chiral anomaly is crucial transport evidence for TSMs. In order to obtain a conclusive and intrinsic NLMR, it is recommended to take the measurement at a high field exceeding $B_Q$, which pushes the chemical potential into the lowest Landau level [24]. The $B_Q$ might be considerably high for the TaP family due to their high carrier densities ($10^{18} \sim 10^{19}$ cm$^{-3}$) [35,40]. On the other hand, the ultrahigh mobility gives rise to a relatively low onset field of current jetting effect ($B_c$). Such a condition ($B_Q \gg B_c$) makes the measurement of NLMR really difficult as the current jetting effect takes effect at low fields. However, a giant PHE or AMR is observed in TaP, which comes mainly from the increased $\rho_\perp$ instead of the reduced $\rho_\parallel$. The large anisotropic orbital magnetoresistance is responsible for the giant PHE.

The current jetting effect not only deforms the PHE and AMR curves for the increased $\mu B$, but also leads to a double-peak structure in the AMR curves for appropriate $\mu B$. We classify these two effects of field-induced non-uniform current distribution into classical and Hall-angle-type current jetting effect, respectively. Nevertheless, the angular dependence of PHE curves ($\rho_{yx}$) does not change (i.e., period in $\pi$ and maximums at $\pm 45°$ and $\pm 135°$ remain) and the altered profile is restored when $\mu B$ is reduced. Our work lays the giant PHE/AMR and the large current jetting in TaP on the same base, i.e., the ultrahigh mobility. Although the former suggests promising applications in spintronics, the latter shows the other side of the coin, which needs to be considered in the future device design.

## ACKNOWLEDGMENTS

This work was supported by the National Key R&D Program of China (Grant Nos. 2016YFA0300404, 2017YFA0403600 and 2017YFA0403502), the National Natural Science Foundation of China (Grant Nos. U1532267, 11674327, 11874363, 51603207, 11574288, U1732273 and 11504379), and the Natural Science Foundation of Anhui Province (Grant No. 1708085MA08). J.Y. and W.L.Z. contributed equally to this work.

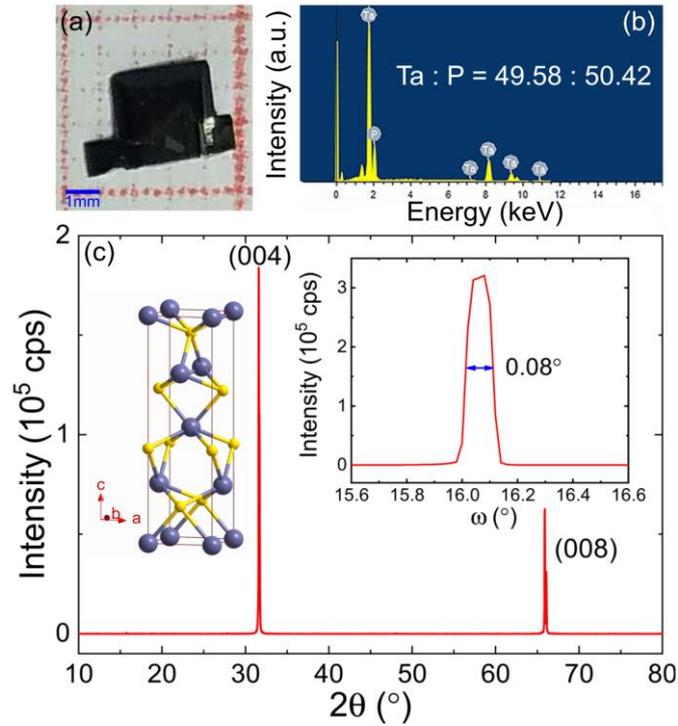

FIG.1. (a) Optical image of as-grown TaP single crystal. (b) EDS result of single crystal, with atomic ratio shown in the spectroscopy. (c) Single crystal XRD pattern taken at room temperature. Left inset: crystal structure of TaP. Blue and yellow spheres represent Ta and P atoms, respectively. Right inset: rocking curve scan of the [004] diffraction, showing a narrow FWHM=0.08 °.

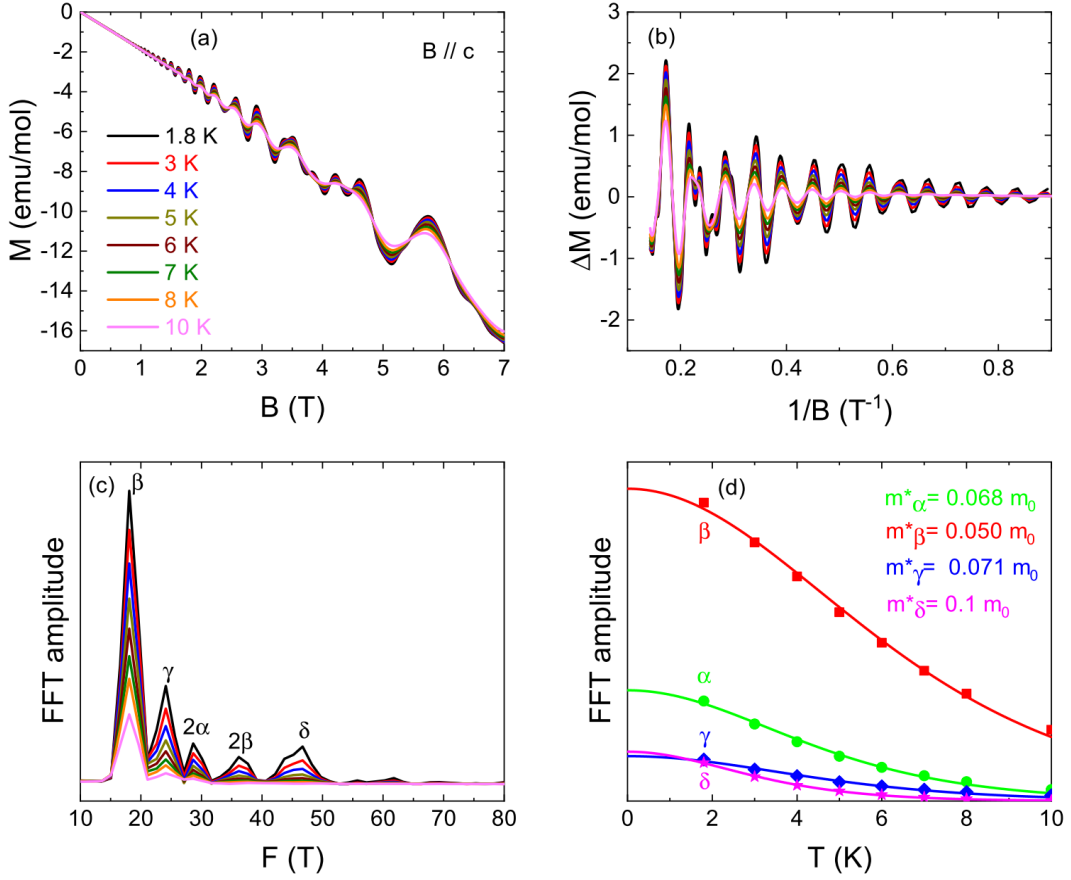

FIG. 2. (a) Magnetization as a function of magnetic field ($B \parallel c$) for TaP taken at various temperatures. (b) Oscillatory part of magnetization obtained after removing the background. (c) FFT spectra of the dHvA oscillations in (b). (d) Temperature-dependent FFT amplitudes of $F_\alpha$, $F_\beta$, $F_\gamma$ and $F_\delta$, fitted to the temperature damping factor $R_T$ of the LK formula.

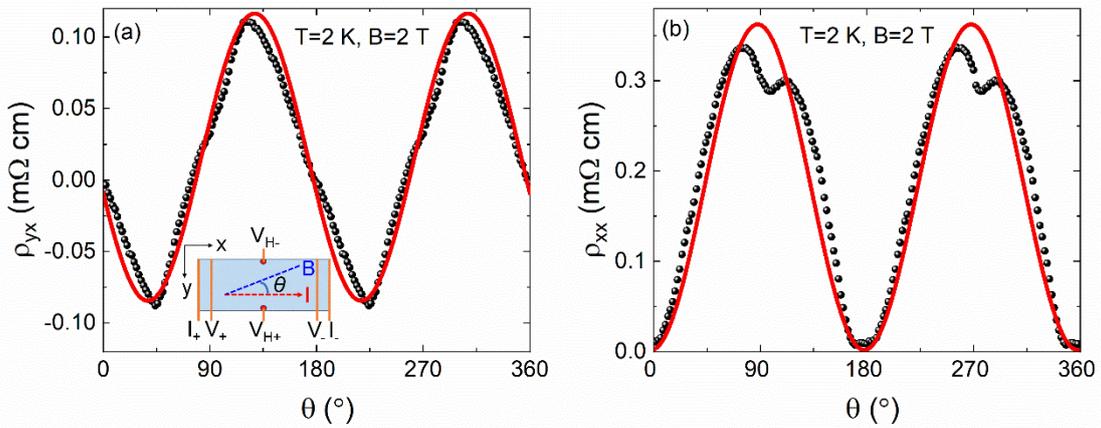

FIG. 3. Angular dependence of (a) planar Hall resistivity $\rho_{yx}$ and (b) longitudinal MR

$\rho_{xx}$ for TaP taken at 2 K and 2 T. Red solid curves represent the fits to Eqs. (1) and (2), respectively. Inset in (a): schematic diagram of measurement geometry.

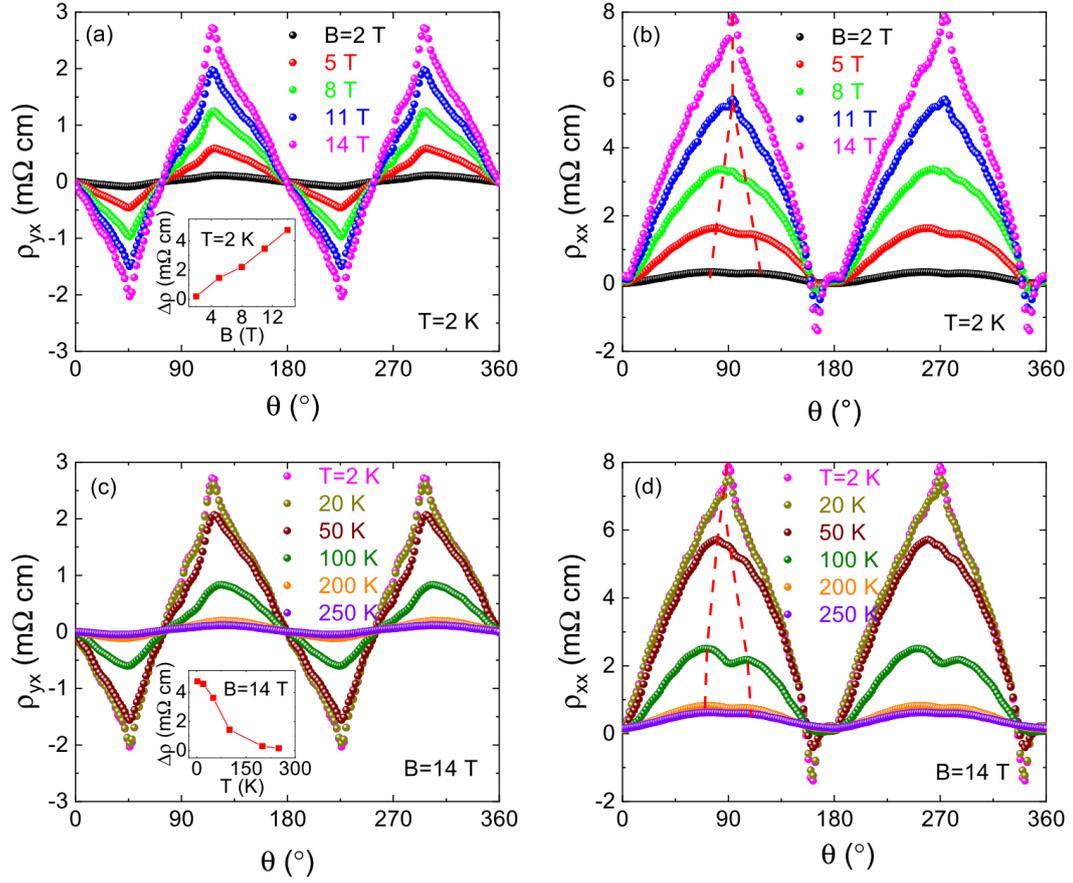

FIG. 4. Angular dependence of (a) planar Hall resistivity $\rho_{yx}$ and (b) longitudinal MR $\rho_{xx}$ for TaP taken at 2 K and various magnetic fields, and angular dependence of (c) $\rho_{yx}$ and (d) $\rho_{xx}$ taken at various temperatures and 14 T. Insets of (a) and (c): chiral anomaly induced anisotropic resistivity $\Delta\rho$ as a function of magnetic field and temperature, respectively. Dashed lines in (b) and (d) indicate the double peaks of $\rho_{xx}$ around 90 °.

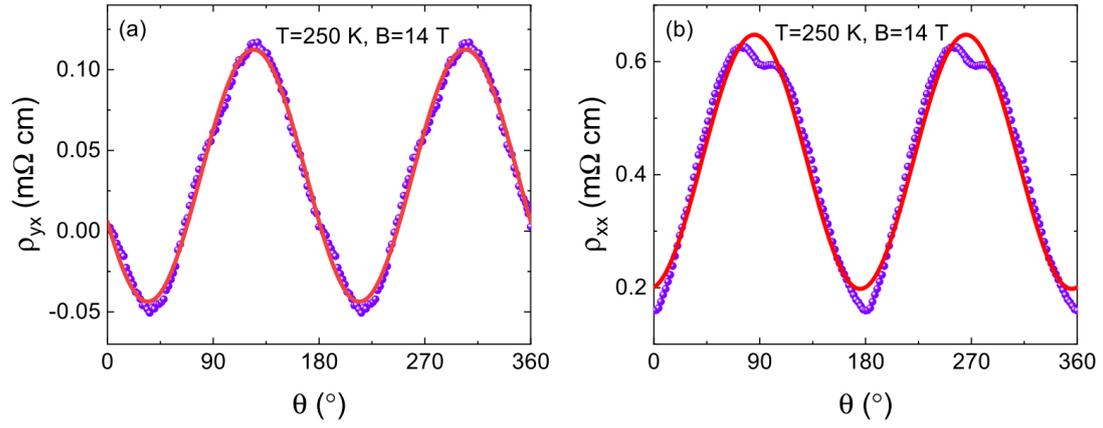

FIG. 5. Angular dependence of (a) planar Hall resistivity $\rho_{yx}$ and (b) longitudinal MR $\rho_{xx}$ for TaP taken at 250 K and 14 T. Red solid curves represent the fits to Eqs. (1) and (2), respectively.